\documentclass{PoS}

\usepackage{feynmp}
\usepackage{cite}
\usepackage{xspace}
\usepackage{multirow}
\usepackage{textcomp}

\newcommand{\mpost}[1]{\immediate\write18{mpost "#1"}}

\newcommand{\ket}[1]{\ensuremath{|#1\rangle}\xspace}
\newcommand{\bra}[1]{\ensuremath{\langle #1|}\xspace}

\newcommand{\elemm}[3]{\ensuremath{\bra{#1}{#2}\ket{#3}}\xspace}

\title{Lattice study of $\pi\pi$ scattering using $\mathrm{N}_\mathrm{f}\mathrm{=2+1}$ Wilson improved quarks with masses down to their physical values}

\ShortTitle{$\pi\pi$ scattering in the $\rho$ channel down to the physical point.}

\author{\speaker{Thibaut Metivet}, on behalf of the Budapest-Marseille-Wuppertal collaboration\\
        CEA Saclay, IRFU, SPhN\\
        E-mail: \email{thibaut.metivet@cea.fr}}

\abstract
{
We use 2HEX smeared gauge configurations generated with an $\mathrm{N}_\mathrm{f}\mathrm{=2+1}$ clover improved Wilson action to investigate $\pi\pi$ scattering in the $\rho$ channel. The range of lattice spacings (0.054 to 0.12 fm) and space-like extents (32 and 48) allows us to extract the scattering parameters through the volume dependence of the $\pi\pi$-state energies according to L\"uscher's formalism. The pion masses (134 to 300 MeV) are light enough to allow the decay of the rho and the level repulsion observed indicates that our data are sensitive to the interaction. We analyse our data with a multi-channel GEVP variational formula. Our results are in good agreement with the experimental values and consistent with a weak pion mass dependence of the $\rho\pi\pi$ coupling constant.
}

\FullConference{The 32nd International Symposium on Lattice Field Theory,\\
		23-28 June, 2014\\
		Columbia University New York, NY}

\begin{document}

\section{Introduction}
Most lattice QCD calculations performed nowadays include dynamical sea-quarks effects, which allows the spontaneous emergence of quark-antiquark pairs from the vacuum and therefore opens the way to resonant states. As light quark masses are reduced, these states find enough phase space to materialize and decay, thus providing profound insight into nonperturbative aspects of QCD.

The $\rho$ meson is an emblematic example of such resonant hadron state which decays nearly exclusively into two pions. The finite-volume formalism to analyse scattering states on the lattice was developed by M. L\"uscher \cite{Luscher:1991cf, Luscher:1986pf, Luscher:1990ux} and we use it to compute the $\rho$ resonance parameters.

The study is carried out with gauge configurations which extend down to the physical pion mass \cite{Durr:2010vn, Durr:2010aw}.

\section{$\pi\pi$ scattering in finite volume}

We consider a system of two $\pi$ mesons with zero total momentum and $I=J=1$, in a cubic box of size $L$ with periodic boundary conditions. The boundary conditions impose a quantization of the momenta: $\vec{k}=(2\pi/L)\vec{n}$ with $\vec{n}\in\mathbb{Z}^3$. The pions are back-to-back with momenta $\pm\vec{k}\neq 0$ since $J=1$.

In the absence of interactions, the energies of the system follow the dispersion relation $E_n=2\:\sqrt{(2\pi/L)^2\vec{n}^2+m_{\pi}^2}$. Switching interactions on leads to two different kinds of finite-volume effects. The first are due to self-energy corrections to the individual pions and are suppressed exponentially with $m_{\pi} \, L$; they are negligible for $m_{\pi} \, L \gtrsim 4$. The second type of finite-volume corrections is due to scattering. As they result from short-range interactions between the particles, these corrections are expected to be of order $1/L^3$ and directly related to the scattering amplitude.

In finite volume, the energies of the two-particle state are shifted from their free field values $E_n=2\:\sqrt{(2\pi/L)^2 n^2+m_{\pi}^2}$ by the interaction and the quantization condition can be expressed through L\"uscher's formula \cite{Luscher:1986pf,Luscher:1991cf}:
\begin{equation}
\cot \delta (q) = \frac{1}{q \, \pi^{3/2}}\, \mathcal{Z}_{00}(1;q^{2})
\end{equation}
with $q=(L/2\pi)\sqrt{E^2/4-m_{\pi}^2}$ the reduced momentum, $\delta$ the $I=J=1$ scattering phase-shift, and $$\mathcal{Z}_{00}(s;q^{2})=\frac{1}{\sqrt{4\pi}}\sum_{\vec{n}\in\mathbb{Z}^{3}}{(n^2-q^2)^{-s}}$$ the generalized zeta function analytically continued in the complex plane. Efficient ways to compute $\mathcal{Z}_{00}(1;q^{2})$ numerically can be found in \cite{Luscher:1990ux} and \cite{Lellouch:2011qw}.

\begin{figure}
\centering
\includegraphics[scale=0.3,angle=-90]{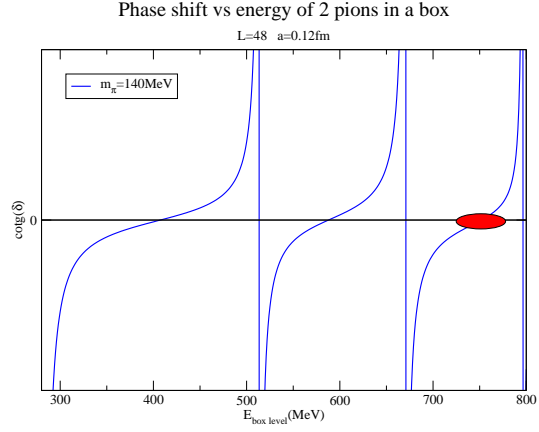}
\caption{Plots of $\cot \delta(E)$ computed with L\"uscher's formula for $m_{\pi}=140 \textrm{ MeV}$. The red spot shows the physical $\rho$ resonance region. The singularities of $\cot \delta(E)$ correspond to non-interacting $2\pi$ states. The lowest energy considered is the minimal energy of two back-to-back pions with $J=1$ in the box $E_{min}=2\:\sqrt{(2\pi/L)^2+m_{\pi}^2}$.}
\label{fig:phase shift vs energy of 2pi in the box}
\end{figure}

L\"uscher's formula allows us to compute the $\pi\pi$ phase shifts at several discrete momenta from a determination of the spectrum. As we are interested in the $\rho$ channel, it is reasonable to focus on the physical $\rho$ energy region, around $780$ MeV. As shown on Fig.\ref{fig:phase shift vs energy of 2pi in the box}, this region corresponds to \emph{excited} $2\pi$ states and require the use of more advanced extraction techniques such as the variational method of \cite{Luscher:1990ck}.

Note that L\"uscher's formula is valid as long as one stays below the $4\pi$ inelastic threshold. However, we assume that the highly suppressed $2\pi\to 2n\pi \;(n>1)$ processes in the $\rho$ channel can be neglected, which allows us to carry the analysis at low -- and even physical -- pion masses.

\section{Methodology}

\subsection{Excited states extraction}
In order to obtain information about $\pi\pi$ scattering in the $\rho$ resonance region, one may need to extract excited states energies from the correlation functions computed on the lattice. To this end, we use the variational method of \cite{Luscher:1990ck}, which consists in using a cross correlator matrix $C_{ij}(t)$ instead of the usual single correlator:
\begin{equation}
C_{ij}(t)=\langle \mathcal{O}_{i}(t)\: \bar{\mathcal{O}_{j}}(0) \rangle
\end{equation}
with $\{\mathcal{O}_{i}\}$ a set of $N$ independent appropriate interpolators.
$C_{ij}(t)$ has the usual spectral decomposition
\begin{equation}
C_{ij}(t)=\sum_n^\infty \elemm{0}{\mathcal{O}_{i}(0)}{n} \elemm{n}{\mathcal{O}_{j}^{\dag}(0)}{0} e^{-E_n t}
\end{equation}
which can be written in matrix form
\begin{equation}
C(t)=V D(t) V^{\dag}
\end{equation}
with $V_{i\,n}=\elemm{0}{\mathcal{O}_{i}(0)}{n}$ and $D(t)=\textrm{diag}(e^{-E_n t})$.

Now, the main idea behind the variational method is that, at large time, one can neglect the contributions coming from high-energy states because of the exponential decay. In that case $V$ and $D$ are finite matrices and one has:
\begin{equation}
C(t)C^{-1}(t_0)=VD(t)V^{\dag} (VD(t_0)V^{\dag})^{-1}=V D(t-t_0) V^{-1}
\end{equation}
so that the eigenvalues of $C(t)C^{-1}(t_0)$ -- solutions of the generalized eigenvalue problem $C(t)u=\lambda(t,t_0)C(t_0)u$ -- provide the requested energies through:
\begin{equation}
\lambda_n(t,t_0) = e^{-E_n (t-t_0)}
\end{equation}

In practice, we diagonalize $C(t)C^{-1}(t_0)$ for each $t$ and fixed $t_0$ to obtain the generalized eigenvalues and then extract the energies from the plateaus of $\log\lambda_n(t,t_0)/(t_0-t)$ at large $t$. Note that computation times limit the number of independent interpolators that we can use, and $t$ and $t_0$ must be large enough to minimize the contamination of higher levels. A more comprehensive analysis of these contaminations can be found in \cite{Blossier:2009kd}.

\subsection{Operators, inversions and contractions}
In our implementation, we use up to 5 independent operators with the quantum numbers of the $\rho$. The first operator is a 2-quark "$\rho$" operator:
\begin{equation}
\mathcal{O}_{\rho} = \bar{u}\:\gamma_i\: u - \bar{d}\:\gamma_i\: d
\end{equation}
with $\gamma_i$ an euclidean Dirac matrix.
The other independent operators are built from non-local $I=1$ combinations of charged $\pi$ interpolators with different lattice momenta:
\begin{equation}
\mathcal{O}_{\pi\pi}(t,\vec{p}) = \pi^{+}(\vec{p})\pi^{-}(-\vec{p}) - \pi^{-}(\vec{p})\pi^{+}(-\vec{p})
\end{equation}
with $\vec{p}=(2\pi/L)\vec{n}$ and $\pi^{\pm}(\vec{p})=\sum_{\vec{x}}\bar{q}(\vec{x})\gamma_5 q'(\vec{x}) e^{i\vec{p}\cdot\vec{x}}$, $q$, $q'=u \textrm{ or }d$.

\begin{figure}[h]
\centering
\def\mydiagramwidth{12}
\def\mydiagramheight{20}
\def\mydiagramblack{\fmfv{decor.shape=circle,decor.filled=full,decor.size=2thick}}
\def\mydiagramgray{\fmfv{decor.shape=circle,decor.filled=gray50,decor.size=2thick}}
\def\mydiagramarrow{2mm}
\begin{fmffile}{fgraphs}
\vspace{0.5cm}
\begin{fmfgraph*}(\mydiagramwidth,\mydiagramheight)
\fmfset{arrow_len}{\mydiagramarrow}
\fmfbottom{i1,i2}
\fmftop{o1,o2}
\fmf{fermion,left=0.3}{i1,o1,i1}
\fmf{fermion,left=0.3}{i2,o2,i2}
\mydiagramgray{i1,i2}
\mydiagramblack{o1,o2}
\fmflabel{$\vec{p}$}{i1}
\fmflabel{$\vec{q}$}{i2}
\fmflabel{$-\vec{p}$}{o1}
\fmflabel{$-\vec{q}$}{o2}
\end{fmfgraph*}
\hfill{\huge -}\hfill
\begin{fmfgraph*}(\mydiagramwidth,\mydiagramheight)
\fmfset{arrow_len}{\mydiagramarrow}
\fmfbottom{i1,i2}
\fmftop{o1,o2}
\fmf{fermion,left=0.3}{i1,o2,i1}
\fmf{fermion,left=0.3}{i2,o1,i2}
\mydiagramgray{i1,i2}
\mydiagramblack{o1,o2}
\fmflabel{$\vec{p}$}{i1}
\fmflabel{$\vec{q}$}{i2}
\fmflabel{$-\vec{p}$}{o1}
\fmflabel{$-\vec{q}$}{o2}
\end{fmfgraph*}
\hfill{\huge+} \hfill
\begin{fmfgraph*}(\mydiagramwidth,\mydiagramheight)
\fmfset{arrow_len}{\mydiagramarrow}
\fmfbottom{i1,i2}
\fmftop{o1,o2}
\fmf{fermion}{i1,i2,o1,o2,i1}
\mydiagramgray{i1}
\mydiagramblack{o1}
\fmflabel{$\vec{p}$}{i1}
\fmflabel{$\vec{q}$}{i2}
\fmflabel{$-\vec{p}$}{o1}
\fmflabel{$-\vec{q}$}{o2}
\end{fmfgraph*}
\hfill{\huge+} \hfill
\begin{fmfgraph*}(\mydiagramwidth,\mydiagramheight)
\fmfset{arrow_len}{\mydiagramarrow}
\fmfbottom{i1,i2}
\fmftop{o1,o2}
\fmf{fermion}{i1,o2,o1,i2,i1}
\mydiagramgray{i1}
\mydiagramblack{o1}
\fmflabel{$\vec{p}$}{i1}
\fmflabel{$\vec{q}$}{i2}
\fmflabel{$-\vec{p}$}{o1}
\fmflabel{$-\vec{q}$}{o2}
\end{fmfgraph*}
\hfill{\huge-}\hfill
\begin{fmfgraph*}(\mydiagramwidth,\mydiagramheight)
\fmfset{arrow_len}{\mydiagramarrow}
\fmfbottom{i1,i2}
\fmftop{o1,o2}
\fmf{fermion}{i1,i2,o2,o1,i1}
\mydiagramgray{i2}
\mydiagramblack{o1}
\fmflabel{$\vec{p}$}{i1}
\fmflabel{$\vec{q}$}{i2}
\fmflabel{$-\vec{p}$}{o1}
\fmflabel{$-\vec{q}$}{o2}
\end{fmfgraph*}
\hfill{\huge-}\hfill
\begin{fmfgraph*}(\mydiagramwidth,\mydiagramheight)
\fmfset{arrow_len}{\mydiagramarrow}
\fmfbottom{i1,i2}
\fmftop{o1,o2}
\fmf{fermion}{i1,o1,o2,i2,i1}
\mydiagramgray{i2}
\mydiagramblack{o1}
\fmflabel{$\vec{p}$}{i1}
\fmflabel{$\vec{q}$}{i2}
\fmflabel{$-\vec{p}$}{o1}
\fmflabel{$-\vec{q}$}{o2}
\end{fmfgraph*}
\\\vspace{1.5cm}
\begin{fmfgraph*}(\mydiagramwidth,\mydiagramheight)
\fmfset{arrow_len}{\mydiagramarrow}
\fmfbottom{i1,i2}
\fmftop{o}
\fmf{fermion}{i1,i2,o,i1}
\mydiagramgray{i2}
\mydiagramblack{o}
\fmflabel{$\vec{p}$}{i1}
\fmflabel{$\vec{q}$}{i2}
\fmflabel{$-\vec{P}$}{o}
\end{fmfgraph*}
\hfill{\huge-}\hfill
\begin{fmfgraph*}(\mydiagramwidth,\mydiagramheight)
\fmfset{arrow_len}{\mydiagramarrow}
\fmfbottom{i1,i2}
\fmftop{o}
\fmf{fermion}{i1,o,i2,i1}
\mydiagramgray{i2}
\mydiagramblack{o}
\fmflabel{$\vec{p}$}{i1}
\fmflabel{$\vec{q}$}{i2}
\fmflabel{$-\vec{P}$}{o}
\end{fmfgraph*}
\hspace{8cm}
\begin{fmfgraph*}(\mydiagramwidth,\mydiagramheight)
\fmfset{arrow_len}{\mydiagramarrow}
\fmfbottom{i}
\fmftop{o1,o2}
\fmf{fermion}{i,o1,o2,i}
\mydiagramgray{i}
\mydiagramblack{o1}
\fmflabel{$-\vec{p}$}{o1}
\fmflabel{$-\vec{q}$}{o2}
\fmflabel{$\vec{P}$}{i}
\end{fmfgraph*}
\hfill{\huge-}\hfill
\begin{fmfgraph*}(\mydiagramwidth,\mydiagramheight)
\fmfset{arrow_len}{\mydiagramarrow}
\fmfbottom{i}
\fmftop{o1,o2}
\fmf{fermion}{i,o2,o1,i}
\mydiagramgray{i}
\mydiagramblack{o1}
\fmflabel{$-\vec{p}$}{o1}
\fmflabel{$-\vec{q}$}{o2}
\fmflabel{$\vec{P}$}{i}
\end{fmfgraph*}
\\\vspace{0.5cm}
\end{fmffile}
\caption{The contractions of $\pi\pi\to\pi\pi$ (top), $\pi\pi\to\rho$ (bottom-left) and $\rho\to\pi\pi$ (bottom-right), the $\rho\to\rho$ being trivial. Time flows upward from $0$ to $t$. Black dots represent an explicit summation whereas shaded dots represent a noise-noise contact. Between those dots we can have stochastic (one-line segments) or generalized stochastic (two-line segments) propagators. Reprinted from \cite{Frison:2010ws}.}
\label{fig:contractions}
\end{figure}
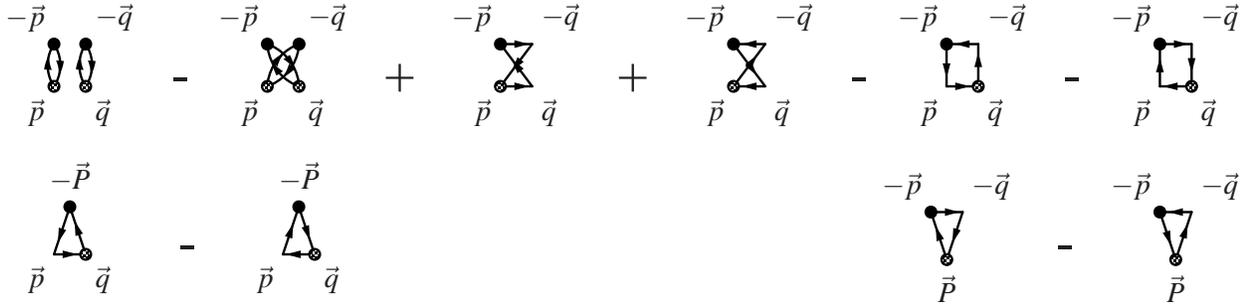

When used to build the cross correlator, this type of operator requires all-to-all propagators as one wants to fix the momentum of each pion independently. We use stochastic propagators and stochastic generalized propagators to compute the contractions, following \cite{Aoki:2007rd}.
The different contractions are shown in Fig.\ref{fig:contractions}.

\mpost{fgraphs}

\subsection{Parametrization of the resonance}
In order to extract the phase shift in the region of interest from the few discrete values computed on the lattice with L\"uscher's formula, we assume that $\pi\pi$ scattering is dominated by a narrow $\rho$ resonance, which we describe as a Breit-Wigner:
\begin{equation}
\sin^2\delta\simeq \frac{\Gamma_\rho^2}{4(E-M_\rho)^2+\Gamma_\rho^2}
\end{equation}
with $M_\rho$, $\Gamma_\rho$ the mass and width of the resonance.

In addition, to eliminate the strong dependence of $\Gamma_\rho$ on the kinematics and hence on the quark masses, we use an effective interaction lagrangian:
\begin{equation}
\mathcal{L}_{int}=g_{\rho\pi\pi}\; \epsilon_{abc} \, \rho^a_\mu \, \pi^b \partial^\mu \pi^c
\end{equation}
which gives at tree level
\begin{equation}
\Gamma_\rho=\frac{g_{\rho\pi\pi}^2}{6\pi}\:\frac{k_\rho^3}{M_\rho^2}
\end{equation}
with $k_\rho=\sqrt{M_\rho^2/4-m_{\pi}^2}$. The coupling $g_{\rho\pi\pi}$ may furthermore have a small dependence on the pion mass, providing a convenient parametrization of the width.

Inserted into L\"uscher's formula, this leads to:
\begin{equation}
M_\rho^2 = E^2 + g_{\rho\pi\pi}^2 \:\frac{4\sqrt{\pi}}{3} \:\frac{q^2 \mathcal{Z}_{00}(1;q^{2})}{EL^3}
\end{equation}
where $M_\rho$ and $g_{\rho\pi\pi}$ are the unknowns we want to determine from measurements of at least two $E$'s. Since we always have at least two energy levels from the variational method, this formula can be used either directly in a system of equations involving two levels, or in a fit with all the levels. For the latter, we actually compute and fit $\sin^2\delta(E)$ using the equivalent formula:
\begin{equation}
\sin^2\delta(E)
=\left(1 + \frac{\mathcal{Z}_{00}^2(1;q^{2})}{q^2 \pi^3}\right)^{-1}
=\left(1+ \left[ \frac{6\pi}{g_{\rho\pi\pi}^2}\frac{E}{k^3}(M_\rho^2-E^2) \right]^2\right)^{-1}
\label{eq:sin2d(E)}
\end{equation}

\section{Results}
We use the Budapest-Marseille-Wuppertal collaboration setup \cite{Durr:2008zz, Durr:2008rw} with a tree-level $O(a^2)$-improved Symanzik action, tree-level $O(a)$-improved Wilson fermions, $N_f=2+1$ flavors and 2 steps of HEX gauge-link smearing \cite{Durr:2010vn, Durr:2010aw, Kurth:2010yk}. We present the preliminary results obtained for 6 independent gauge ensembles, with properties summarized in Table \ref{table:ensembles}. The errors are purely statistical and are computed with 2000 bootstraps. The analysis has been refined since the oral presentation and we now extract the $\rho$ resonance parameters with a fit of $\sin^2\delta$ versus $E$ when possible, dropping the few ($\sim 20 \%$) bootstrap samples with a fit relative error greater than $100\%$.

\begin{table}[h]
\centering
\begin{tabular}{|lcrcrccc|}
\hline
$\;\beta$ & $am_{ud}^\mathrm{bare}$ & $am_s^\mathrm{bare}$ & volume & \#\,traj.\ & $am_\pi$ & $m_\pi L$ & $t_0$\\
\hline
\multirow{3}{*}{3.31}
 & -0.09300 & -0.0400 & $32^3\times 48$ & 2500 & 0.1771(05) & 5.65 & 5 \\
 & -0.09756 & -0.0400 & $32^3\times 48$ & 2600 & 0.1202(11) & 4.00 & 5 \\
 & -0.09933 & -0.0400 & $48^3\times 48$ & 1240 & 0.0804(13) & 3.94 & 8 \\
\hline
\multirow{3}{*}{3.61}
 & -0.03121 &  0.0045 & $48^3\times 48$ & 2200 & 0.1211(2) & 3.87 & 10 \\
 & -0.03300 &  0.0045 & $48^3\times 48$ & 2100 & 0.1026(4) & 4.93 & 8 \\
 & -0.03440 &  0.0045 & $48^3\times 48$ & 1100 & 0.0864(4) & 4.15 & 8 \\
\hline
\end{tabular}
\caption{\label{table:ensembles}\sl
Overview of our $N_f\!=\!2\!+\!1$ simulations.
The scales at $\beta=3.31,3.61$
are $a^{-1}=1.697(6),2.561(26)$ GeV,
respectively. The last column gives the $t_0$ that we use in the variational method.}
\end{table}

\begin{figure}
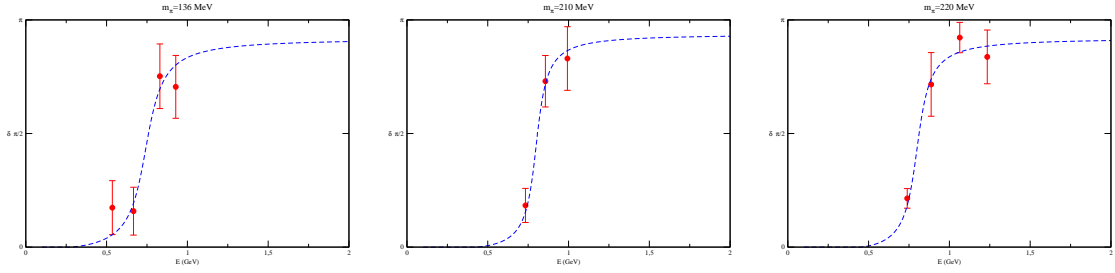


\begin{minipage}{.33\textwidth}
\centering
\includegraphics[scale=0.20]{delta_vs_E_with_fit_136}
\end{minipage}
\begin{minipage}{.33\textwidth}
\centering
\includegraphics[scale=0.20]{delta_vs_E_with_fit_210}
\end{minipage}
\begin{minipage}{.33\textwidth}
\centering
\includegraphics[scale=0.20]{delta_vs_E_with_fit_220}
\end{minipage}

\caption{\label{fig:sin2d_vs_E_with_fit}Preliminary results for the $\mathrm{I=J=1}$ partial wave phase shifts computed on the lattice for different pion masses. The dashed blue curves represent the fits to Eq. (\protect\ref{eq:sin2d(E)}).}
\end{figure}

Fig.\ref{fig:sin2d_vs_E_with_fit} shows the values of $\delta(E)$ vs $E$ for ensembles where we have extracted more than 2 states as well as the corresponding fits.

\begin{figure}[!h]
\begin{minipage}{.5\textwidth}
\centering
\includegraphics[scale=0.27,angle=-90]{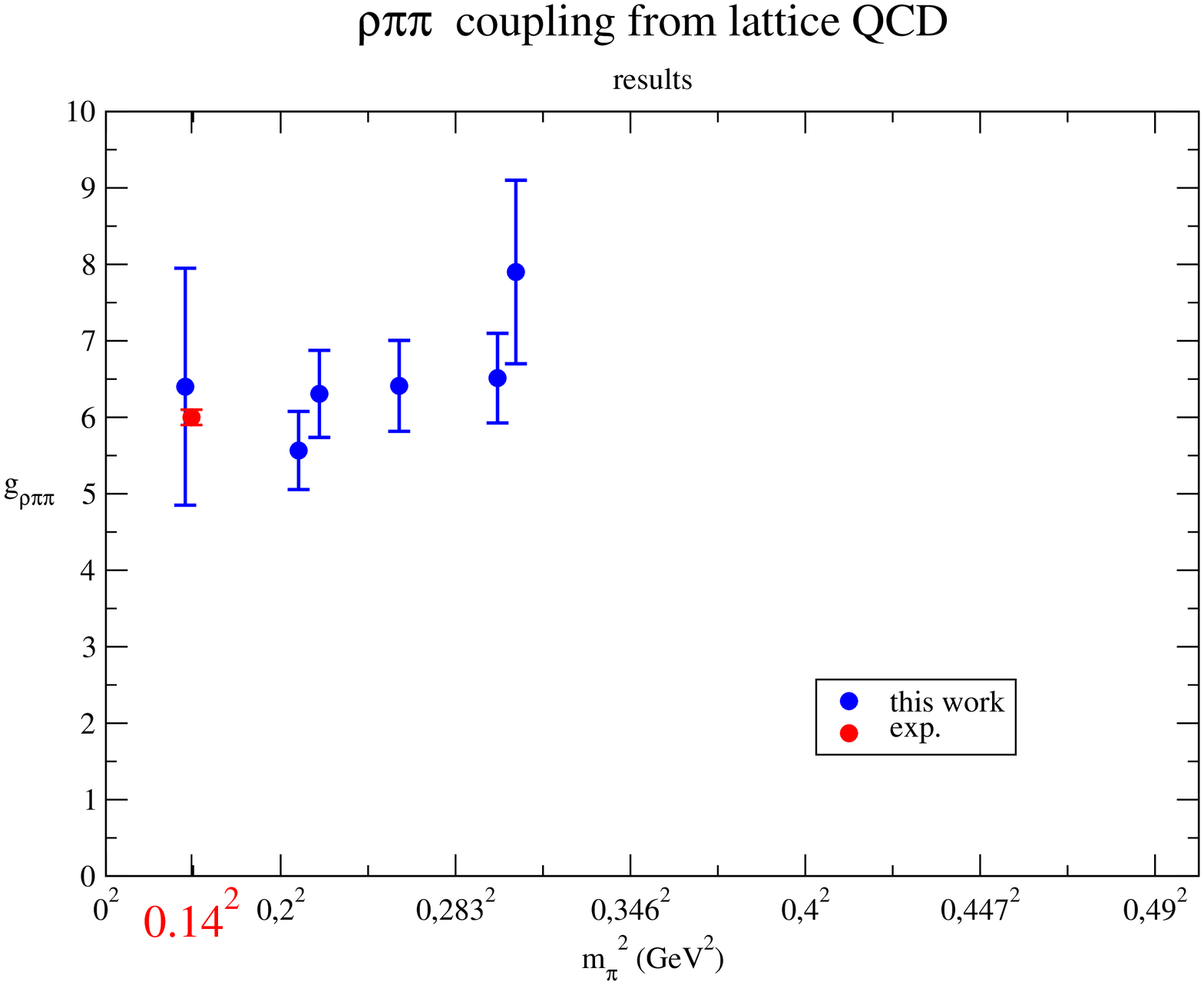}
\end{minipage}
\begin{minipage}{.5\textwidth}
\centering
\includegraphics[scale=0.27,angle=-90]{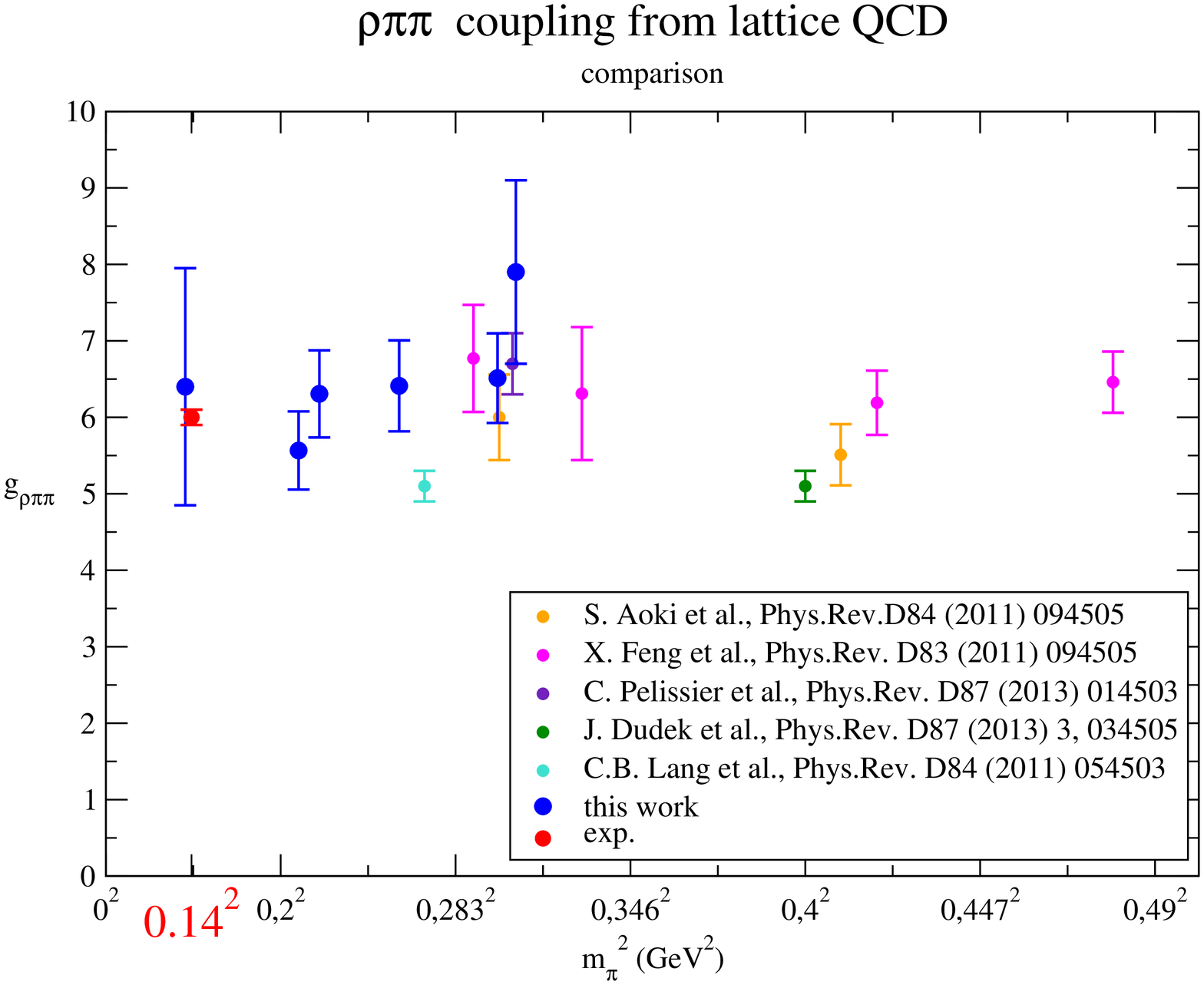}
\end{minipage}

\begin{minipage}{.5\textwidth}
\centering
\includegraphics[scale=0.27,angle=-90]{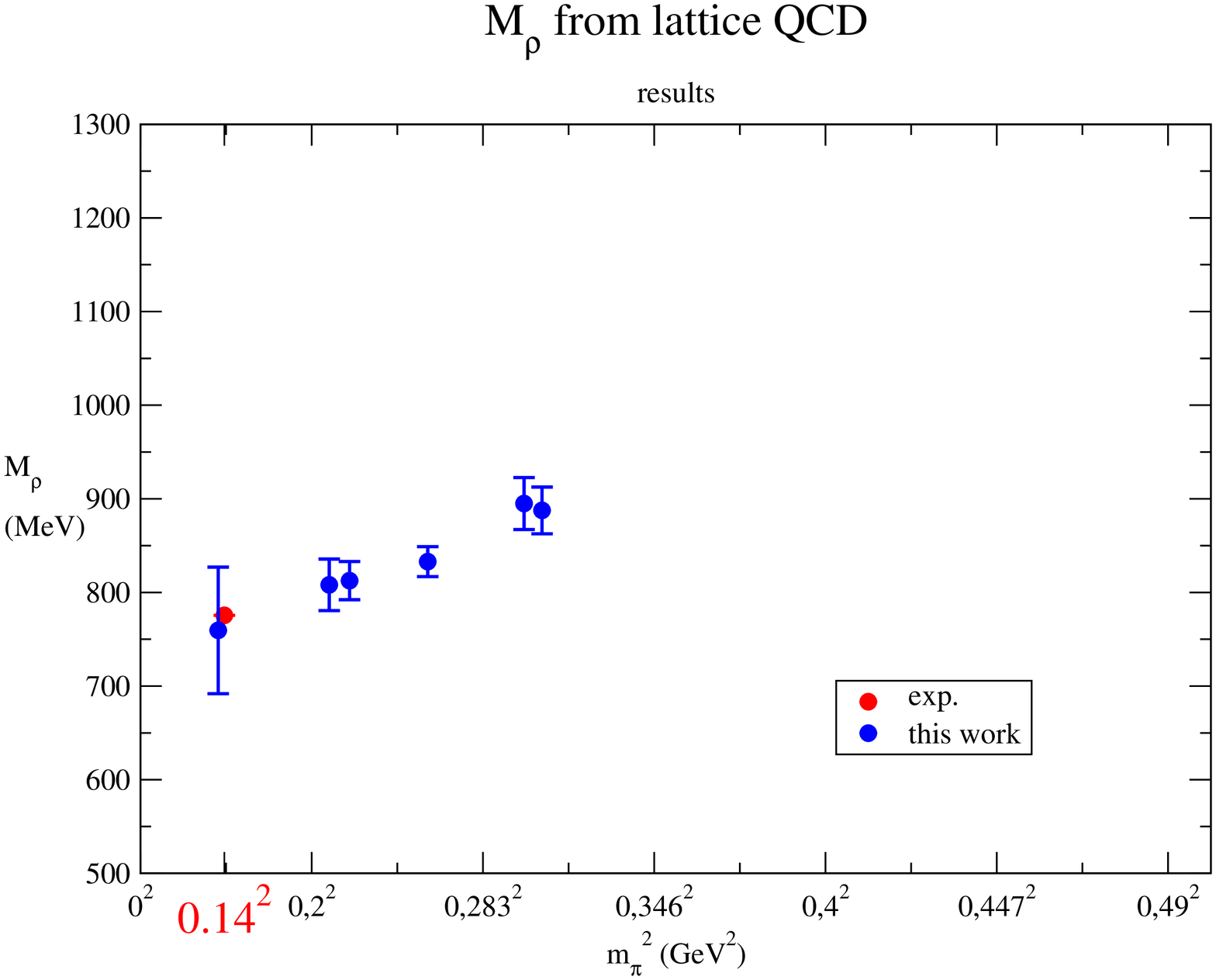}
\end{minipage}
\begin{minipage}{.5\textwidth}
\centering
\includegraphics[scale=0.27,angle=-90]{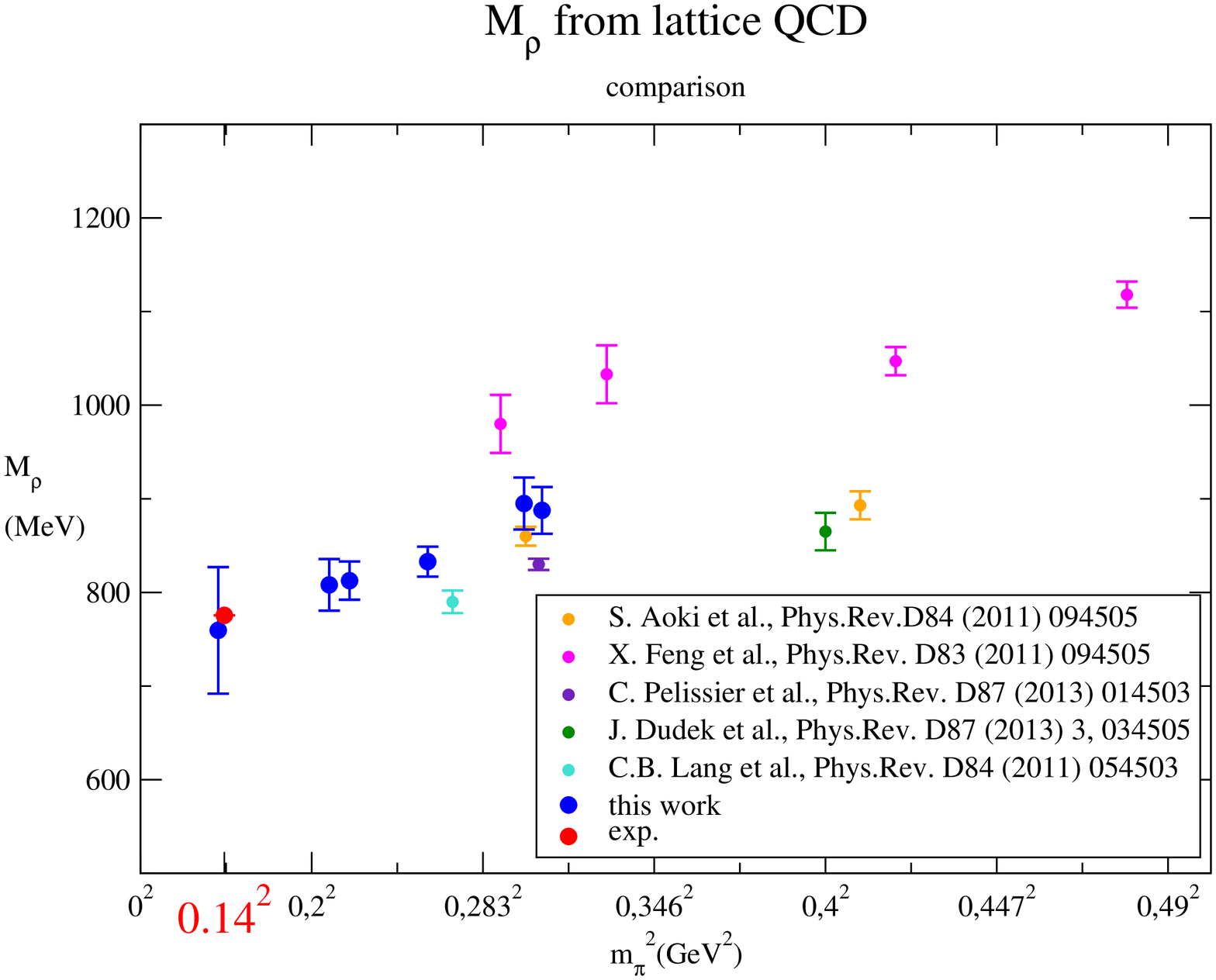}
\end{minipage}
\caption{\label{fig:g and Mrho results}Results for $g_{\rho\pi\pi}$ and $M_\rho$, with purely statistical errors.}
\end{figure}

Our preliminary results for $g_{\rho\pi\pi}$ and $M_\rho$ vs $m_{\pi}^2$ are presented in Fig.\ref{fig:g and Mrho results}, together with the results obtained by other collaborations. These results confirm the weak dependence of $g_{\rho\pi\pi}$ on the pion mass, and a constant fit gives at the physical point:
\begin{equation}
g_{\rho\pi\pi}^{phys}=6.2 \pm 0.24
\end{equation}
which is in good agreement with the experimental value. 

We are currently finalizing our analysis. Though very challenging, it would also be interesting to combine the methods studied here with those developed for lattice QCD+QED in \cite{Borsanyi:2014jba}, to help shed some light on the controversies surrounding isospin breaking effects in the $\rho$ resonance parameters \cite{Agashe:2014kda}.

\section*{Aknowledgments}
Computations were performed using HPC resources provided by
GENCI-[IDRIS] (grant 52275) and FZ J\"ulich. This work was supported
in part by the OCEVU Labex (ANR-11-LABX-0060) and the A$^\star$MIDEX
project (ANR-11-IDEX-0001-02), funded by the "Investissements d'Avenir"
French government program and managed by the ANR, by CNRS grants GDR
$n^o$2921 and PICS $n^o$4707, by EU grants FP7/2007-2013/ERC 208740
and MRTN-CT-2006-035482 (FLAVIAnet), and by DFG grants FO 502/2, SFB-TR
55.

\bibliographystyle{hieeetr}
\bibliography{pos_metivet}

\begin{thebibliography}{10}

\bibitem{Luscher:1991cf}
M.~Luscher, ``{Signatures of unstable particles in finite volume},'' {\em
  Nucl.Phys.}, vol.~B364, pp.~237--254, 1991.

\bibitem{Luscher:1986pf}
M.~Luscher, ``{Volume Dependence of the Energy Spectrum in Massive Quantum
  Field Theories. 2. Scattering States},'' {\em Commun.Math.Phys.}, vol.~105,
  pp.~153--188, 1986.

\bibitem{Luscher:1990ux}
M.~Luscher, ``{Two particle states on a torus and their relation to the
  scattering matrix},'' {\em Nucl.Phys.}, vol.~B354, pp.~531--578, 1991.

\bibitem{Durr:2010vn}
S.~Durr, Z.~Fodor, C.~Hoelbling, S.~Katz, S.~Krieg, {\em et~al.}, ``{Lattice
  QCD at the physical point: light quark masses},'' {\em Phys.Lett.},
  vol.~B701, pp.~265--268, 2011.

\bibitem{Durr:2010aw}
S.~Durr, Z.~Fodor, C.~Hoelbling, S.~Katz, S.~Krieg, {\em et~al.}, ``{Lattice
  QCD at the physical point: Simulation and analysis details},'' {\em JHEP},
  vol.~1108, p.~148, 2011.

\bibitem{Lellouch:2011qw}
L.~Lellouch, ``{Flavor physics and lattice quantum chromodynamics},'' {\em Les
  Houches 2009, Session XCIII}, pp.~629--698, 2011.

\bibitem{Luscher:1990ck}
M.~Luscher and U.~Wolff, ``{How to Calculate the Elastic Scattering Matrix in
  Two-dimensional Quantum Field Theories by Numerical Simulation},'' {\em
  Nucl.Phys.}, vol.~B339, pp.~222--252, 1990.

\bibitem{Blossier:2009kd}
B.~Blossier, M.~Della~Morte, G.~von Hippel, T.~Mendes, and R.~Sommer, ``{On the
  generalized eigenvalue method for energies and matrix elements in lattice
  field theory},'' {\em JHEP}, vol.~0904, p.~094, 2009.

\bibitem{Aoki:2007rd}
S.~Aoki {\em et~al.}, ``{Lattice QCD Calculation of the rho Meson Decay
  Width},'' {\em Phys.Rev.}, vol.~D76, p.~094506, 2007.

\bibitem{Frison:2010ws}
J.~Frison {\em et~al.}, ``{Rho decay width from the lattice},'' {\em PoS},
  vol.~LATTICE2010, p.~139, 2010.

\bibitem{Durr:2008zz}
S.~Durr, Z.~Fodor, J.~Frison, C.~Hoelbling, R.~Hoffmann, {\em et~al.},
  ``{Ab-Initio Determination of Light Hadron Masses},'' {\em Science},
  vol.~322, pp.~1224--1227, 2008.

\bibitem{Durr:2008rw}
S.~Durr, Z.~Fodor, C.~Hoelbling, R.~Hoffmann, S.~Katz, {\em et~al.}, ``{Scaling
  study of dynamical smeared-link clover fermions},'' {\em Phys.Rev.},
  vol.~D79, p.~014501, 2009.

\bibitem{Kurth:2010yk}
T.~Kurth {\em et~al.}, ``{Scaling study for 2 HEX smeared fermions: hadron and
  quark masses},'' {\em PoS}, vol.~LATTICE2010, p.~232, 2010.

\bibitem{Borsanyi:2014jba}
S.~Borsanyi, S.~Durr, Z.~Fodor, C.~Hoelbling, S.~Katz, {\em et~al.}, ``{Ab
  initio calculation of the neutron-proton mass difference},'' 2014,
  arXiv:hep-lat/1406.4088.

\bibitem{Agashe:2014kda}
K.~Olive {\em et~al.}, ``{Review of Particle Physics},'' {\em Chin.Phys.},
  vol.~C38, p.~090001, 2014.

\end{thebibliography}

\end{document}